\documentclass[12pt]{article}

\usepackage{amsmath,amssymb,amsfonts,amsthm}
\usepackage{graphicx}
\usepackage{cite}
\usepackage[all]{xy}
\usepackage[toc,page]{appendix}
\usepackage{hyperref}
\hypersetup{colorlinks=true,  citecolor=red, linkcolor=blue}

\allowdisplaybreaks[3]

\newcounter{propositiona}

\newcounter{definitiona}
\newcommand{\definitiona}[1]{\refstepcounter{definitiona}
\noindent
\textbf{Definition \thedefinitiona.}\, #1}
\newcounter{remarka}
\newcommand{\remarka}[1]{\refstepcounter{remarka}
\noindent
\textbf{Remark \theremarka.}\, #1}
\newcounter{examplea}
\newcommand{\examplea}[1]{\refstepcounter{examplea}
\noindent
\textbf{Example \theexamplea.}\, #1}
\newcounter{lemmaa}

\newcounter{theorema}
\newcommand{\theorema}[1]{\refstepcounter{theorema}
\noindent
\textbf{Theorem\, \thetheorema.}\, {\it #1}}
\newcounter{corollarya}

\def\beq{\begin{equation}}
\def\eeq{\end{equation}}
\def\barr{\begin{array}{ll}}
\def\earr{\end{array}}

\def\vec#1{{\boldsymbol{\rm #1}}} 

\def\Div{\mathop{\hbox{\rm Div}}}
\def\Curl{\mathop{\hbox{\rm Curl}}}

\title{Invariant Reduction for Partial Differential Equations. I: Conservation Laws and Systems with Two Independent Variables}

\author{ 
Kostya Druzhkov\footnotemark[1],~~Alexei Cheviakov\footnotemark[2]\vspace{0.5cm}\\
\small $^{\rm a,b}$\emph{Department of Mathematics and Statistics, University of Saskatchewan, Saskatoon, Canada}\vspace{0.2cm}\\
}

\begin{document}

\footnotetext[1]{Corresponding author. Electronic mail: konstantin.druzhkov@gmail.com}
\footnotetext[2]{Electronic mail: shevyakov@math.usask.ca}

\maketitle \numberwithin{equation}{section}

\begin{abstract}

For a system of partial differential equations that has an extended Kovalevskaya form, a reduction procedure is presented that allows one to use a local (point, contact, or higher) symmetry of a system and a symmetry-invariant conservation law to algorithmically calculate constants of motion holding for symmetry-invariant solutions. Several examples including cases of point and higher symmetry invariance are presented and discussed. An implementation of the algorithm in \verb|Maple| is provided.

\end{abstract}


\section{Introduction}

The set of conservation laws admitted by a system of partial differential equations (PDE) contains essential information about that system. In particular, conservation laws describe rates of change of physical quantities, express differential constraints, provide divergence forms of PDEs required for existence, uniqueness, stability, and global solution behaviour analysis, and allow for the introduction of nonlocal variables. Conservation laws are also used for seeking exact solutions of PDEs through potential systems and the Fokas method. Symmetry-invariant conservation laws are used for double reduction to seek symmetry-invariant exact solutions of PDE models more efficiently.  Conservation laws are broadly used in numerical solvers requiring divergence forms of governing equations, such as finite element and finite volume methods; conservation law-preserving finite difference discretizations can be systematically constructed. For details and applications, see Refs.~\cite{BCAbook,BCA, kelbin2013new, deconinck2014method, colbrook2019unified, T1972, johnson2016orbital, knops1984quasiconvexity, batchelor2000introduction, frisch1995turbulence, AncoChev2020, Olver, VinKr, marsden1994mathematical, popovych2010conservation, leveque1992numerical, cheviakov2020invariant} and references therein.

While in general, conserved quantities can be described by various classes of differential forms, in practice, one often seeks local conservation laws represented by differential $(n-1)$-forms (on a given PDE system, understood as a geometric object) that are, loosely speaking, closed on solutions, where $n$ is the number of independent variables for the system. They can be described using divergence-type relations holding on solutions.
In particular, in the (1+1)-dimensional $(t, x)$-space, a local conservation law is determined by a divergence-type relation
\beq\label{eq:intro:CL:2d}
D_t (\widetilde{P}_1) + D_x (\widetilde{P}_2) =0
\eeq
between differential functions $\widetilde{P}_1$, $\widetilde{P}_2$, or equivalently, by a differential $1$-form
$$
\widetilde{\omega} = \widetilde{P}_1 dx - \widetilde{P}_2\hspace{0.15ex} dt\,,
$$
where the differential $d\hspace{0.15ex} \widetilde{\omega}$ vanishes on solutions. The conservation law is represented by the restriction $\omega$ of $\widetilde{\omega}$ to the PDE system.

In three dimensions $(x,y,z)$, conservation laws are determined by expressions of the form
\beq\label{eq:intro:CL:nd}
\Div {\widetilde{\vec P}} = D_i \widetilde{P}^i =0
\eeq
where the summation in $i$ taken from $1$ to $n=3$ is assumed, $D_i$ are total derivative operators\footnote{See Section \ref{Sec2} for details.}, and $\Div$  denotes the total divergence. Another type of conserved quantity in three dimensions corresponds to classes of $(n-2)$-forms closed on solutions. It is given by a curl-type expression
\beq\label{eq:intro:gCL:nd}
\Curl {\widetilde{\vec Q}} = 0
\eeq
in terms of some differential functions ${\widetilde{\vec Q}}$ and the total curl operator.

It is important to note that conservation laws can be systematically constructed using the application of Euler operators to their characteristic forms (see, e.g., Refs.~\cite{anco2002direct, anco2002direct2, BCAbook}). It is also essential that conservation laws are invariant with respect to coordinate transformations. In particular, a divergence-type expression is mapped by a coordinate transformation into a divergence-type expression. Most systems arising in applications are $\ell$-normal~(see Ref.~\cite{VinKr} for details). For example, all systems in an extended Kovalevskaya form belong to this class, as well as all Lagrangian non-gauge systems. For an $\ell$-normal system, the non-triviality of conservation laws is verified using their cosymmetries~\cite{VinKr}, i.e., restrictions of their characteristics to the system. Each non-trivial conservation law of such a system corresponds to a unique non-trivial cosymmetry. The converse may not be true, since cosymmetries may correspond to Lagrangians admissible by differential equations. Cosymmetries of an $\ell$-normal system can be found using symbolic computations~\cite{KraVerVit}.

A practically useful way of construction of exact solutions to nonlinear and linear PDE systems is the calculation of solutions invariant with respect to a given Lie point or contact symmetry. Through the use of canonical local coordinates, a symmetry reduction leads to a system with fewer independent variables; in particular, a PDE system with two independent variables leads to an ODE system which is generally easier to solve. This often yields a useful explicit subset of solutions of the given PDE system. It is also possible to seek solutions invariant with respect to higher symmetries; the components of the corresponding evolutionary fields vanish for invariant solutions, which leads to an overdetermined PDE system that invariant solutions satisfy.

When a conservation law is invariant under the action of a symmetry of a PDE system, it provides additional information about the structure of invariant solutions. In two dimensions, it yields a constant of motion. A common approach in the case of point or contact symmetry group invariance is the use of canonical local coordinates.
However, in the cases when such computations are impractical or technically prohibitive, as well as in the cases of higher symmetries that do not yield a group of transformations (a \emph{flow}) and hence yield no canonical coordinates, the above technique cannot be applied. Besides, in general, canonical coordinates provide only local information about invariant solutions.

\medskip\noindent\textbf{The Main Idea.} The current paper is concerned with PDEs and PDE systems with two independent variables. The main idea of the paper is described in Section~\ref{MainIdea}. In a nutshell, because local conservation laws are equivalence classes of differential forms, symmetries act on them by means of the Lie derivative. This observation is true for point, contact, and higher symmetries, despite the fact that the latter generate no geometric flows. If $E_{\varphi}$ is a symmetry in the evolutionary form, then it vanishes on $E_{\varphi}$-invariant solutions. Hence the Lie derivative of a differential form with respect to $E_{\varphi}$ vanishes on the invariant solutions.

\noindent
\textbf{The Principal Result} (see Theorem~\ref{maintheor}). If $E_{\varphi}$ is a symmetry in the evolutionary form and a differential $1$-form $\omega$ represents an invariant conservation law of a $2$-dimensional PDE system, then there is a function $\vartheta$ such that the restrictions of $\mathcal{L}_{E_{\varphi}} \omega$ and $d\vartheta$ to any solution of this system coincide. Such a function $\vartheta$ is constant on any $E_{\varphi}$-invariant solution.

\bigskip

For PDEs with two independent variables, invariant solutions satisfy finite-dimensional systems. In this sense they are similar to solutions of ODEs. Their constants of motion play the same role as first integrals for ODEs. In particular, these constants of motion can be useful in the study of global qualitative properties of invariant solutions. The knowledge of a sufficient number of constants of motion allows one to obtain a general solution in exactly the same way as in the case of ODEs.
Importantly, in the course of computations of constants of invariant motion, no changes of local coordinates are required. The algorithm is simple and holds for both point/contact and higher symmetries, with or without the known Lie group form. The algorithm is implemented in symbolic software in a straightforward way; such an implementation for one of the examples in the paper is contained in Appendix \ref{App:A}.

This paper generalizes the reduction mechanism proposed in Refs.~\cite{sjoberg2007double, sjoberg2009double} and developed in Refs.~\cite{bokhari2010generalization, anco2020symmetry} (see also references therein). We consider invariant conservation laws in a coordinate-free manner. This allows us to cover the case of higher symmetries. It is also worth mentioning another reduction mechanism introduced in Ref.~\cite{anderson1997symmetry}; the latter approach allows one to use some groups of transformations in the multidimensional case without requiring their solvability. However, the use of group-invariant representatives is essential for that method.

The paper is organized as follows. In Section \ref{Sec2}, an algorithm for computing constants of invariant motion is presented, with the running example of the Burgers equation illustrating every step. Section \ref{SecEx} contains additional examples: the KdV equation, a potential system of the Kaup-Boussinesq equations, and a potential Boussinesq system.

The paper is concluded with a Discussion section. Appendix A contains an example of implementation of the algorithm presented in this work in \verb|Maple| symbolic software for the Kaup-Boussinesq potential system. The code can be generalized to other systems in a straightforward way.

We use the Einstein summation notation throughout this paper.

\section{From conservation laws to constants of invariant motion}\label{Sec2}

Let us consider a system of $m$ evolution equations
\begin{align}
\begin{aligned}
&u^1_t = f^1,\\
&\quad \ldots\,,\\
&u^m_t = f^m,
\end{aligned}
\label{EvoSys}
\end{align}
where $f^1$, $\ldots, f^m$ are functions of the two independent variables $t$, $x$, dependent variables $u^1, \ldots, u^m$ and a finite number of $x$-derivatives $u^i_{x}$, $u^i_{xx}$, $\ldots$ ($i = 1, \ldots, m$).
The total derivative operators are given by the formulae
\begin{align*}
D_t = \partial_t + u^i_t\partial_{u^i} + u^i_{tx}\partial_{u^i_x} + u^i_{tt}\partial_{u^i_t} + \ldots\,,\qquad
D_x = \partial_x + u^i_x\partial_{u^i} + u^i_{xx}\partial_{u^i_x} + u^i_{tx}\partial_{u^i_t} + \ldots\,.
\end{align*}
We assume that they are prolonged to derivatives of all orders.

\vspace{1ex}

\remarka{
Any system of equations in an extended Kovalevskaya form
\begin{align*}
u^i_{k_i t} = \tilde{f}^i
\end{align*}
can be rewritten as a system of evolution equations through an introduction of auxiliary variables of the form $v^i = u^i_t$, $\ldots$. Here for each $i$, $k_i\geqslant 1$ is an integer,  $u^i_{k_i t}$ is the $k_i$-th order $t$-derivative of $u^i$ (no summation is implied here), and none of the functions $\tilde{f}^j$ depend on $u^i_{k_i t}$ or its derivatives. For example, the equation $u_{ttt} = u_{ttxx}$ transforms into the evolution system
\begin{align*}
u_t = v\,,\qquad v_t = w\,,\qquad w_t = w_{xx}\,.
\end{align*}
Let us note that most systems of differential equations that arise in applications can be written in an extended Kovalevskaya form using coordinate changes. For instance, the Benjamin-Bona-Mahony equation
\begin{align*}
u_t + u_x + uu_x - u_{txx} = 0
\end{align*}
can be written in an extended Kovalevskaya form
\begin{align*}
u_{\hat{t}\hat{t}\hat{t}} = \tilde{f}(u, u_{\hat{t}}, u_{\hat{x}}, u_{\hat{t}\hat{t}\hat{x}}, u_{\hat{t}\hat{x}\hat{x}}, u_{\hat{x}\hat{x}\hat{x}})
\end{align*}
(and hence, as an evolution system) after the transformation $\hat{t} = t + x$, $\hat{x} = t - x$.}

\vspace{1ex}

Let $\mathcal{E}$ be system~\eqref{EvoSys} together with all of its differential consequences:
\begin{align*}
\mathcal{E}\, \colon\qquad  u^i_t = f^i\,,\quad u^i_{tt} = D_t(f^i)\,,\quad u^i_{tx} = D_x(f^i)\,,\quad u^i_{ttt} = D_t^{\hspace{0.1ex} 2}(f^i)\,,\quad \ldots\qquad (i = 1, \ldots, m)\,.
\end{align*}
Here $D_t^{\hspace{0.1ex} 2} = D_t\circ D_t$, etc. Denote by $\mathcal{F}(\mathcal{E})$ the algebra of $C^{\infty}$-functions of $t$, $x$, $u^1, \ldots, u^m$ and a finite number of $x$-derivatives $u^i_{x}$, $u^i_{xx}$, $\ldots$ If a smooth function $r$ depends on $u^1_t, \ldots, u^m_t$ or their derivatives, one can eliminate these variables using equations of the system $\mathcal{E}$. We denote by $r|_{\mathcal{E}}$ the result of such an elimination.

Let us introduce the restriction $\,\overline{\!D}_t$ of the operator $D_t$ to $\mathcal{E}$,
\begin{align*}
\,\overline{\!D}_t = \partial_t + f^i\partial_{u^i} + D_x(f^i)\partial_{u^i_x} + D^2_x(f^i)\partial_{u^i_{xx}} + D^3_x(f^i)\partial_{u^i_{xxx}} + \ldots
\end{align*}
For functions $\varphi^1, \ldots, \varphi^m\in\mathcal{F}(\mathcal{E})$, we denote by $E_{\varphi}$ the corresponding evolutionary vector field
\begin{align*}
E_{\varphi} = \varphi^i\partial_{u^i} + D_t(\varphi^i)\partial_{u^i_t} + D_x(\varphi^i)\partial_{u^i_x} + D_t^2(\varphi^i)\partial_{u^i_{tt}} + \ldots
\end{align*}
and call the vector function $\varphi = (\varphi^1, \ldots, \varphi^m)^T$ its \textit{characteristic}.

\subsection{Evolutionary symmetries}

\vspace{1ex}

\definitiona{An (infinitesimal, evolutionary) \textit{symmetry} of system~\eqref{EvoSys}} is an evolutionary vector field $E_{\varphi}$ such that
\begin{align*}
E_{\varphi}(u^i_t - f^i)|_{\mathcal{E}} = 0 \qquad \text{for}\quad i = 1, \ldots, m\,.
\end{align*}

\vspace{1ex}
\noindent
Let us recall that each generator
\begin{align*}
Y = \xi^t(t, x, u^1, \ldots, u^m)\partial_{t} + \xi^x(t, x, u^1, \ldots, u^m)\partial_{x} + \eta^i(t, x, u^1, \ldots, u^m)\partial_{u^i}
\end{align*}
of a point symmetry for system~\eqref{EvoSys} gives rise to its evolutionary symmetry having the characteristic with components $\varphi^i = \eta^i - f^i\xi^t - u^i_{x}\xi^x$.
The same applies to contact symmetries. In this work, we consider any evolutionary symmetries
\begin{align*}
E_{\varphi} = \varphi^i\partial_{u^i} + \ldots,
\end{align*}
including higher symmetries where the components $\varphi^i$ may depend on $x$-derivatives of $u^j$ of arbitrarily high order.

\bigskip \emph{Throughout Section~\ref{Sec2}, we illustrate various concepts using the same example of the Burgers equation.}

First, it is instructive to demonstrate our approach using an evolutionary symmetry arising from a point symmetry generator.

\bigskip

\examplea{\label{BurgSymexample} Consider the Burgers equation
\begin{align}
u_t = uu_x + u_{xx}\,,
\label{Burgers}
\end{align}
which is of the form \eqref{EvoSys} with $u^1 = u$, $f^1 = uu_x + u_{xx}$. The characteristic
$$
\varphi = x + tu + t^2(uu_x + u_{xx}) + txu_x
$$
gives rise to its evolutionary symmetry
\begin{align*}
&E_{\varphi} = (x + tu + t^2(uu_x + u_{xx}) + txu_x)\partial_u + D_t(x + tu + t^2(uu_x + u_{xx}) + txu_x)\partial_{u_t} + \ldots
\end{align*}
Here $E_{\varphi}(u_t - uu_x - u_{xx})$ vanishes on the corresponding system $\mathcal{E}$ since
\begin{align*}
E_{\varphi}(u_t - uu_x &- u_{xx}) = D_t(\varphi) - u_x\varphi - uD_x(\varphi) - D_x^2(\varphi) ={}\\
&= (t + t^2u_x) (u_t - uu_x - u_{xx}) + (tx + t^2u)D_x(u_t - uu_x - u_{xx}) + t^2D_x^2(u_t - uu_x - u_{xx})\,.
\end{align*}
This evolutionary symmetry arises from the point symmetry generator
\beq\label{eq:Burg:symm:Y}
Y = -t^2\partial_t - tx\partial_x + (x + tu)\partial_u\,.
\eeq

\vspace{1ex}

\remarka{One can describe the local flow generated by $Y$ as follows:
\begin{align}
t' = \dfrac{t}{1 + \tau t}\,,\qquad x' = \dfrac{x}{1 + \tau t}\,,\qquad u' = u + \tau (x + tu)\,,
\label{flow}
\end{align}
where $\tau$ is the group parameter. However, in general, the approach we propose \textit{is not based on flows} of symmetries. Moreover, higher symmetries often do not generate flows even at the level of the intrinsic geometry of differential equations.
}
}

\subsection{Conservation laws}

A conservation law of system~\eqref{EvoSys} in its global form is given by a relation
\begin{align}
\int_{x_1}^{x_2} P_1 dx\,\bigg|_{t_1}^{t_2} = \int_{t_1}^{t_2} P_2\hspace{0.15ex} dt\, \bigg|_{x_1}^{x_2}
\label{conslaw}
\end{align}
that must hold on every smooth solution of~\eqref{EvoSys} for any rectangle $\Pi = [t_1; t_2]\times [x_1; x_2]$ that lies in its domain. It shows that on a segment $[x_1; x_2]$, the total value of the quantity having density $P_1$ can change over time only due to the boundary flux given by $P_2$. In particular, this means that there are no sources of the conserved density $P_1$ within a finite interval $[x_1; x_2]$.
Without loss of generality, we assume that $P_1, P_2\in \mathcal{F}(\mathcal{E})$, because on solutions, $u^i_t = f^i$, $u^i_{tt} = D_t(f^i)$, etc.

The relation~\eqref{conslaw} can be written in the form
\begin{align*}
\int_{\partial \Pi} P_1 dx - P_2\hspace{0.15ex} dt = 0.
\end{align*}
Then the Stokes' theorem implies that on solutions of system~\eqref{EvoSys},
\begin{align*}
\iint_{\Pi} \big(D_t(P_1) + D_x(P_2)\big) dt\wedge dx = 0.
\end{align*}
Here it is reasonable to require the triviality of the differential form $\big(\,\overline{\!D}_t(P_1) + D_x(P_2)\big) dt\wedge dx$. This requirement amounts to the condition $\,\overline{\!D}_t(P_1) + D_x(P_2) = 0$.

If there exists a function $\nu\in \mathcal{F}(\mathcal{E})$ such that $P_1 dx - P_2\hspace{0.15ex} dt = D_x(\nu)dx + \,\overline{\!D}_t(\nu)\hspace{0.2ex} dt$, then on solutions of system~\eqref{EvoSys}, relation~\eqref{conslaw} takes the form
\begin{align*}
\Big(\nu\Big|_{x_1}^{x_2}\Big)\Big|_{t_1}^{t_2} = -\Big(\nu\Big|_{t_1}^{t_2}\Big)\Big|_{x_1}^{x_2}
\end{align*}
We call such conservation laws \emph{trivial}. This discussion motivates the following definition.

\bigskip

\definitiona{\label{Conslawdefin} A \textit{conservation law} of system~\eqref{EvoSys} is an equivalence class of differential forms of the form $P_1 dx - P_2\hspace{0.15ex} dt$ such that $P_1, P_2 \in \mathcal{F}(\mathcal{E})$ and
\begin{align*}
\,\overline{\!D}_t(P_1) + D_x(P_2) = 0.
\end{align*}
Such differential forms are equivalent if they differ by a $1$-form $D_x(\nu)dx + \,\overline{\!D}_t(\nu)\hspace{0.2ex} dt$, where $\nu\in \mathcal{F}(\mathcal{E})$.
}
\vspace{-3ex}
\bigskip

\examplea{\label{BurgCLexample} The Burgers equation~\eqref{Burgers} admits a conservation law represented by the differential form
\begin{align}
u\hspace{0.15ex} dx + \Big(\dfrac{u^2}{2} + u_x\Big) dt\,.
\label{BurgCL}
\end{align}
Here $P_1 = u$, $P_2 = -\Big(\dfrac{u^2}{2} + u_x\Big)$, and $\,\overline{\!D}_t(P_1) + D_x(P_2) = 0$. An equivalent differential form
\begin{align*}
u\hspace{0.15ex} dx + \Big(\dfrac{u^2}{2} + u_x\Big) dt + D_x(u)\hspace{0.2ex} dx + \,\overline{\!D}_t(u)\hspace{0.2ex} dt = (u + u_x)\hspace{0.2ex} dx + \Big(\dfrac{u^2}{2} + u_x + uu_x + u_{xx}\Big) dt
\end{align*}
represents the same conservation law.
}

\bigskip

\remarka{Conservation laws of non-stationary PDEs can be considered analogs of first integrals of ODEs in the following sense. In analytical mechanics, one assigns numbers that do not change over time to instantaneous states of mechanical systems. In this case, instantaneous states are modeled by points (of manifolds called phase spaces), and hence, conservation laws are simply real-valued functions (modulo constants). In continuum mechanics, one can also assign numbers that do not change over time to instantaneous states of mechanical systems. The difference is that here, instantaneous states have a structure of a manifold of dimension $\geqslant 1$. Using appropriate conditions at $x\to \pm \infty$, one can assign the number
$$
\int_{-\infty}^{+\infty} P_1 dx
$$
to an instantaneous state of the entire system located on $(-\infty; +\infty)$. Parts of the mechanical system can exchange parts of this total quantity. Therefore one needs to introduce fluxes to deal with conserved quantities in terms of subsystems of a given mechanical system.
}

\subsection{Conservation laws and cosymmetries}

Definition~\ref{Conslawdefin} reflects the essence of the concept of conservation law, but from the computational point of view, working with equivalence classes is inconvenient. One can deal with conservation laws using their cosymmetries~\cite{VinKr} or their characteristics~\cite{Olver}. For systems of evolution equations, these concepts can be considered equivalent (with some reservations).
In the following, we briefly present some useful facts pertaining to the relation between conservation laws and cosymmetries. Let us note that except for Mart{\'i}nez Alonso lemma, all these facts are given, for example, in Ref.~\cite{vinogradov1984c} in a more general form. We restrict ourselves to special cases relevant to the current consideration.

Each non-trivial conservation law of system~\eqref{EvoSys} defines a unique vector function
$$
\psi = (\psi_1, \ldots, \psi_m)\neq 0
$$
such that $\psi_1, \ldots, \psi_m\in \mathcal{F}(\mathcal{E})$ and the variational derivative of $\psi_i (u^i_t - f^i)$ vanishes on system~\eqref{EvoSys}:
\begin{align}
\dfrac{\delta(\psi_i (u^i_t - f^i))}{\delta u^j}\bigg|_{\mathcal{E}} = 0\quad\text{for}\quad j = 1, \ldots, m.
\label{cosymeq}
\end{align}
We call solutions of equation~\eqref{cosymeq} \textit{cosymmetries}. For evolution equations, the conditions of the Mart\'{i}nez Alonso lemma~\cite{MartinezAlonso} (Lemma 3) are met, and hence a cosymmetry $\psi$ corresponds to a conservation law if and only if the variational derivative vanishes everywhere:
\begin{align}
\dfrac{\delta(\psi_i (u^i_t - f^i))}{\delta u^j} = 0\qquad\quad\text{for}\quad j = 1, \ldots, m.
\label{chareq}
\end{align}
The corresponding conservation law is represented by any differential form $P_1 dx - P_2\hspace{0.2ex} dt$ such that $P_1, P_2 \in \mathcal{F}(\mathcal{E})$ and $\psi_i (u^i_t - f^i) = D_t(P_1) + D_x(P_2)$ (all such forms are equivalent). To find conservation laws of the system~\eqref{EvoSys}, one can solve the equation~\eqref{cosymeq} and then select solutions that also satisfy equation~\eqref{chareq}.

\bigskip

\remarka{A cosymmetry $\psi$ does not correspond to a conservation law of system~\eqref{EvoSys} if and only if the (pre)symplectic operator $l_{\psi} - l_{\psi}^{\hspace{0.2ex} *}$ is non-zero. Here for a characteristic $\chi$ of an arbitrary evolutionary vector field (not necessarily symmetry), the $j$-th component $(l_{\psi} - l_{\psi}^{\hspace{0.2ex} *})(\chi)_j$ of the vector function $(l_{\psi} - l_{\psi}^{\hspace{0.2ex} *})(\chi)$
can be determined from
\begin{align*}
&l_{\psi}(\chi)_j = E_{\chi}(\psi_j) = \sum_{k\geqslant 0} \dfrac{\partial \psi_j}{\partial u^i_{kx}}D_x^{\hspace{0.2ex} k}(\chi^i)\,, \qquad
l_{\psi}^{\hspace{0.2ex} *}(\chi)_j = \sum_{k\geqslant 0} (-1)^k D_x^{\hspace{0.2ex} k} \Big(\dfrac{\partial \psi_i}{\partial u^j_{kx}}\chi^i\Big)\qquad\quad j = 1, \ldots, m\,,
\end{align*}
where $u^i_{kx}$ denotes the $k$-th order $x$-derivative of $u^i$. One can use this condition instead of~\eqref{chareq}. Let us recall that presymplectic operators map characteristics of symmetries of system~\eqref{EvoSys} to its cosymmetries.
}

\bigskip

For a conservation law of~\eqref{EvoSys} represented by a differential form $P_1 dx - P_2\hspace{0.15ex} dt$, the corresponding cosymmetry is given by
\begin{align*}
\psi = \left(\dfrac{\delta P_1}{\delta u^1}, \ldots, \dfrac{\delta P_1}{\delta u^m}\right).
\end{align*}

\vspace{1ex}
\examplea{\label{BurgCosexample} Continuing Example~\ref{BurgCLexample}, we observe that the conservation law represented by differential form~\eqref{BurgCL} corresponds to the cosymmetry $\psi = \delta P_1/\delta u = 1$.}

\subsection{Invariant conservation laws}

Symmetries act on conservation laws by means of the Lie derivative. Thus, if $E_{\varphi}$ is an evolutionary symmetry of~\eqref{EvoSys}, and a differential form $P_1 dx - P_2\hspace{0.15ex} dt$ represents its conservation law, the differential form
\begin{align*}
\mathcal{L}_{E_{\varphi}}(P_1 dx - P_2\hspace{0.1ex} dt) = E_{\varphi}(P_1) dx - E_{\varphi}(P_2)\hspace{0.1ex} dt
\end{align*}
also represents a conservation law of system~\eqref{EvoSys}. It is useful to describe this action in terms of cosymmetries. If $\psi$ is the cosymmetry of a conservation law generated by $P_1 dx - P_2\hspace{0.1ex} dt$, then the cosymmetry of the conservation law generated by $\mathcal{L}_{E_{\varphi}}(P_1 dx - P_2\hspace{0.1ex} dt)$ is $E_{\varphi}(\psi) + l_{\varphi}^{\hspace{0.15ex}*}(\psi)$, where
\begin{align*}
E_{\varphi}(\psi)_j = E_{\varphi}(\psi_j),\qquad
l_{\varphi}^{\hspace{0.15ex}*}(\psi)_j = \sum_{k\geqslant 0} (-1)^k D_x^{\,k} \left(\dfrac{\partial \varphi^i}{\partial u^j_{kx}}\psi_i\right).
\end{align*}
Accordingly, the conservation law with a cosymmetry $\psi$ is $E_{\varphi}$-invariant if and only if
$E_{\varphi}(\psi) + l_{\varphi}^{\hspace{0.15ex}*}(\psi) = 0$. This condition is easy to check. Let us stress that in the general case, this approach allows one to use symmetries (including higher symmetries) to get new conservation laws from known ones.

\vspace{1ex}

\remarka{Cosymmetries of system~\eqref{EvoSys} can be identified with its variational $1$-forms (elements of the group $E^{1,\,n-1}_1(\mathcal{E})$ of the Vinogradov $\mathcal{C}$-spectral sequence~\cite{vinogradov1984c, VinKr}). Symmetries act on variational $1$-forms by means of the Lie derivative. This action can be described in terms of cosymmetries. It is given by the same formula $\psi\mapsto E_{\varphi}(\psi) + l_{\varphi}^{\hspace{0.15ex}*}(\psi)$.}

\bigskip

\examplea{For the symmetry from Example~\ref{BurgSymexample} with the characteristic $\varphi = x + tu + t^2(uu_x + u_{xx}) + txu_x$ and the cosymmetry $\psi = 1$ from Example~\ref{BurgCosexample}, we have
\begin{align*}
E_{\varphi}(\psi) + l_{\varphi}^{\hspace{0.15ex}*}(\psi) &= E_{\varphi}(1) + l_{\varphi}^{\hspace{0.15ex}*}(1) = l_{\varphi}^{\hspace{0.15ex}*}(1) = \sum_{k\geqslant 0} (-1)^k D_x^{\,k} \left(\dfrac{\partial \varphi}{\partial u_{kx}}\right) \\{}
&= \dfrac{\partial \varphi}{\partial u} - D_x\left(\dfrac{\partial \varphi}{\partial u_x}\right) + D_x^2\left(\dfrac{\partial \varphi}{\partial u_{xx}}\right) = t + t^2u_x - D_x(t^2u + tx) - D_x^2(t^2) = 0\,.
\end{align*}
Thus the conservation law represented by differential form~\eqref{BurgCL} is $E_{\varphi}$-invariant.
}

\subsection{Reduction of invariant conservation laws}\label{MainIdea}

One can use symmetries of the system~\eqref{EvoSys} to derive invariant exact solutions. For a symmetry $X = E_{\varphi}$, such invariant solutions are described by the overdetermined system
\begin{align}
u^i_t = f^i\,,\qquad \varphi^i = 0\,,\qquad i = 1, \ldots, m\,.
\label{InvSolSys}
\end{align}
We denote by $\mathcal{E}_X$ the system~\eqref{InvSolSys} with all its differential consequences.
If a conservation law represented by $\omega = P_1 dx - P_2\hspace{0.15ex} dt$ is $X$-invariant, then the Lie derivative $\mathcal{L}_{X} \hspace{0.1ex}\omega$ represents the trivial conservation law, i.e., there is a function $\vartheta\in \mathcal{F}(\mathcal{E})$ such that
\begin{align}
\mathcal{L}_{X} \hspace{0.1ex}\omega = D_x(\vartheta)dx + \,\overline{\!D}_t(\vartheta)dt.
\label{mainformula}
\end{align}
Note that $X$ vanishes on the system $\mathcal{E}_X$. Then the differential form $D_x(\vartheta)dx + \,\overline{\!D}_t(\vartheta)dt$ also vanishes on $\mathcal{E}_X$. Therefore on any $X$-invariant solution, the function $\vartheta$ is constant. We obtain the following

\vspace{1ex}

\theorema{\label{maintheor} Let $X = E_{\varphi}$ be a symmetry of~\eqref{EvoSys}. Suppose that $\omega = P_1 dx - P_2\hspace{0.15ex} dt$ represents an $X$-invariant conservation law of this system, $P_1, P_2\in \mathcal{F}(\mathcal{E})$.
Then a function $\vartheta\in \mathcal{F}(\mathcal{E})$ such that $\mathcal{L}_{X} \hspace{0.1ex}\omega = D_x(\vartheta)dx + \,\overline{\!D}_t(\vartheta)dt$ is constant on any $X$-invariant solution.
}

\vspace{1ex}

Thus the reduction of the conservation law represented by $\omega$ is the constant of $X$-invariant motion $\vartheta$ from~\eqref{mainformula}. Let us emphasize that the choice of a particular conservation law representative $\omega$ plays no role. The restriction $\vartheta|_{\mathcal{E}_X}$ of the function $\vartheta\in \mathcal{F}(\mathcal{E})$ to the invariant surface given by the system $\mathcal{E}_X$ is defined up to an additive constant. In some cases, $\vartheta$ can be trivial, i.e., $\vartheta|_{\mathcal{E}_X}$ can be just a real number (hereinafter we assume, for simplicity, that the surface $\mathcal{E}_X$ is non-singular and connected).

\bigskip

\examplea{Continuing with the Burgers equation, we have $X = E_{\varphi}$, $\varphi = x + tu + t^2(uu_x + u_{xx}) + txu_x$, and $\omega = u\, dx + ({u^2}/{2} + u_x) dt$. Consequently,
\begin{align*}
\mathcal{L}_X \hspace{0.1ex} \omega =&\ (x + tu + t^2(uu_x + u_{xx}) + txu_x)dx {}\\[1ex]
&+ (xu + tu^2 + txuu_x + 1 + 2tu_x + txu_{xx} + t^2(u^2u_x + u_x^2 + 2uu_{xx} + u_{xxx})) dt\,.
\end{align*}
As the corresponding constant of $X$-invariant motion, one can take
\begin{align}
\vartheta = \dfrac{(x + tu)^2}{2} + t(1 + tu_x),
\label{ConstInvBurg}
\end{align}
because $\mathcal{L}_X \hspace{0.1ex} \omega = D_x(\vartheta)dx + \,\overline{\!D}_t(\vartheta)dt$.

\vspace{1ex}
\remarka{We emphasize again that our approach is not based on symmetry transformations. Nevertheless, for illustration purposes, using the flow of $Y$ given by~\eqref{flow}, one can introduce the following canonical local coordinates
\begin{align*}
s = \dfrac{1}{t}\,,\qquad y = \dfrac{x}{t}\,,\qquad w = x + tu\,.
\end{align*}
In these coordinates, the corresponding symmetry generator takes the canonical form $Y = \partial_s$. Regarding $w$ as a dependent variable, one can (locally) rewrite~\eqref{ConstInvBurg} in the form
$$
\vartheta = w^2/2 + w_y\,.
$$
The constant of $X$-invariant motion in the form~\eqref{ConstInvBurg} allows one to conclude that there are no $X$-invariant solutions in a neighborhood of a point $(t_0, x_0)\in \mathbb{R}^2$ with $t_0 = 0$. Indeed, suppose that $u = U(t, x)$ is an $X$-invariant solution. Then there is a real number $c$ such that on the solution domain, we have the identity
\begin{align*}
\dfrac{(x + t\hspace{0.15ex} U)^2}{2} + t\Big(1 + t\,\dfrac{\partial U}{\partial x}\Big) \equiv c\,.
\end{align*}
Substituting $t = 0$, we find
$$
\dfrac{x^2}{2} \equiv c\,.
$$
This identity can not be satisfied on any interval $x\in (x_1, x_2)$.
}
}

\subsection{Formula for constants of invariant motion}

One can derive a general formula for $\vartheta$. From relation~\eqref{mainformula}, it follows that
$$
X(P_1)dx = D_x(\vartheta) dx\,.
$$
Here we can apply the total homotopy formula~\cite{Olver}. The result is defined up to an element of the kernel of the operator $D_x$, i.e., up to an arbitrary function of the variable $t$. It remains to determine this function using~\eqref{mainformula}. Thus, the problem reduces to integration of a known function of $t$.

From the computational point of view, it may be useful to employ \verb|HorizontalHomotopy| function in \verb|Maple| to find $\vartheta$ up to a function of $t$. We provide the horizontal homotopy formula here (up to a function of $t$ and $x$). Namely, there exists a function $h(t, x)$ such that
\begin{align*}
\vartheta = h(t, x) + \int_0^1 \dfrac{G[\tau u]}{\tau} d\tau,\qquad G[u] = \sum_{k=1}^{\infty} \sum_{j=0}^{k-1} (-1)^{\hspace{0.4ex} k - 1 - j}D^{\hspace{0.2ex} k-1-j}_x \Big(\dfrac{\partial (X(P_1))}{\partial u^i_{kx}}\Big) u^i_{jx}\,,
\end{align*}
where $G[u]$ is a differential function of $u$. Substituting this expression for $\vartheta$ to the condition
\begin{align*}
X(P_1) dx - X(P_2)\hspace{0.15ex} dt = D_x(\vartheta)dx + \,\overline{\!D}_t(\vartheta)dt,
\end{align*}
one obtains an equation of the form $dh = g_1(t, x)dx + g_2(t, x)dt$, where $g_i(t, x)$ are known functions. Then, finally,
the constant of $X$-invariant motion has the form
\begin{align}
\vartheta = \int_{0}^xg_1(0; s)ds + \int_0^t g_2(s, x)ds + \int_0^1 \dfrac{G[\tau u]}{\tau} d\tau .
\label{constant}
\end{align}
If $P_1$, $P_2$ and $\varphi$ do not depend on $t$ and $x$, then $g_1 = g_2 = 0$. In this case, a particular form of $P_2$ is not required.

\vspace{1ex}

\examplea{For the Burgers equation, we obtain $X(P_1) = x + tu + t^2(uu_x + u_{xx}) + txu_x$, $G[u] = t^2u^2 + txu + t^2u_x$, and
\begin{align*}
& \vartheta = h(t, x) + \int_0^1 \dfrac{t^2\tau^2 u^2 + tx\tau u + t^2\tau u_x}{\tau}d\tau = h(t, x) + \dfrac{t^2u^2}{2} + txu + t^2u_x\,.
\end{align*}
Substituting this into the relation
\begin{align*}
D_{x}(\vartheta)dx + \,\overline{\!D}_{t}(\vartheta) dt &= (x + tu + t^2(uu_x + u_{xx}) + txu_x)dx +{}\\
&\hspace{1.5ex} + (xu + tu^2 + txuu_x + 1 + 2tu_x + txu_{xx} + t^2(u^2u_x + u_x^2 + 2uu_{xx} + u_{xxx})) dt,
\end{align*}
we get $dh = x dx + dt$, and hence, $\vartheta = (x + tu)^2/2 + t(1 + tu_x)$ is the resulting constant of $X$-invariant motion. It coincides with~\eqref{ConstInvBurg}.
}

\subsection{Constants of invariant motion and cosymmetries}

A cosymmetry of an invariant conservation law plays a role similar to the role of a characteristic for the corresponding constant of invariant motion. Namely, integration by parts shows that there are functions $r_{ki} \in \mathcal{F}(\mathcal{E})$ such that the relation (Noether's identity)
\begin{align}
E_{\chi}(P_1) = \dfrac{\delta P_1}{\delta u^i}\chi^i + D_x(r_{ki} D_x^k(\chi^i))
\label{Noetherident}
\end{align}
holds for all characteristics $\chi$ of evolutionary vector fields. Let $\chi = \varphi$. Because $\delta P_1/\delta u^i = \psi_i$, we get
\begin{align*}
X(P_1) = \psi_i\varphi^i + D_x(r_{ki} D_x^k(\varphi^i))\,.
\end{align*}
Hence one obtains the relation
\begin{align}
\psi_i\varphi^i dx = D_x(\vartheta - r_{ki} D_x^k(\varphi^i))dx\,.
\label{charcons}
\end{align}
Note that $\vartheta - r_{ki} D_x^k(\varphi^i)$ is an equivalent constant of $X$-invariant motion in the sense that its restriction to $\mathcal{E}_X$ coincides with $\vartheta|_{\mathcal{E}_X}$. One can apply the total homotopy formula to~\eqref{charcons}. At this step, the components of conservation laws are not required. But in some cases, it may be inconvenient to check that the derivative $D_t$ of the result vanishes on $\mathcal{E}_X$. The relation~\eqref{mainformula} can be useful here, because $r_{ki} D_x^k(\chi^i)$ can be unambiguously derived using Noether's identity~\eqref{Noetherident}.

\vspace{1ex}

\remarka{The system $\varphi = 0$ can be regarded as a one-parameter family of ODE systems, while $\psi$ can be treated as a family of characteristics of their conservation laws (first integrals). If $\psi$ vanishes on the system
$$
\varphi = 0\,,\qquad D_x(\varphi) = 0\,,\qquad D_x^2(\varphi) = 0\,,\qquad \ldots\,,
$$
then the family of first integrals is a function of $t$. This follows from simple analysis of the Vinogradov $\mathcal{C}$-spectral sequence for ODEs and the relation between cosymmetries and variational $1$-forms. In this case, the restriction of $\vartheta - r_{ki} D_x^k(\varphi^i)$ to $\mathcal{E}_X$ (i.e., $\vartheta|_{\mathcal{E}_X}$) is a function of $t$. In particular, it does not depend on an $X$-invariant solution. But $\vartheta|_{\mathcal{E}_X}$  is a constant of $X$-invariant motion.
Hence $\vartheta|_{\mathcal{E}_X}$ is just a real number, and the constant of $X$-invariant motion is trivial.
}

\section{Examples}\label{SecEx}

\subsection{KdV}

Let us consider the KdV equation
\beq\label{eq:KdV}
u_t = 6uu_x + u_{xxx}
\eeq
and its higher symmetry $X = E_{\varphi}$ with
\beq\label{eq:KdV:5sym}
\varphi = u_{5x} + 10uu_{xxx} + 20u_xu_{xx} + 30u^2u_x\,.
\eeq
The conservation law corresponding to the cosymmetry
\begin{align*}
\psi = c_0 + 2c_1u - c_2(u_{xx} + 3u^2)
\end{align*}
(here $c_0, c_1, c_2\in \mathbb{R}$ are arbitrary) is $X$-invariant, because $X(\psi) + l_{\varphi}^{\hspace{0.2ex} *}(\psi) = 0$. This conservation law is represented by the differential form $P_1 dx - P_2\hspace{0.15ex} dt$, where
\beq\label{eq:KdV:CL}
\barr
P_1 = c_0 u + c_1 u^2 + c_2\left(\dfrac{u_x^2}{2} - u^3\right),{}\\[2ex]
P_2 = -\bigg(c_0 (u_{xx} + 3u^2) + c_1 (4u^3 + 2uu_{xx} - u_x^2) \\[2ex]
\hspace{9ex} \left. +\, c_2\left(u_x u_{xxx} - \dfrac{u_{xx}^2}{2} - 3u^2u_{xx} + 6uu_x^2 - \dfrac{9u^4}{2}\right)\right).
\earr
\eeq
Then $X(P_1) = c_0 \varphi + 2c_1 u \varphi + c_2 (u_x D_x(\varphi)  - 3u^2\varphi)$, and
\begin{align*}
G[u] =\ &c_0(u_{4x} + 20uu_{xx} + 10u_x^2 + 30u^3) + c_1(4uu_{4x} - 4u_xu_{xxx} + 2u_{xx}^2 + 60u^2u_{xx} + 60u^4) +{}\\
&\begin{aligned}
+\, c_2 (2u_x u_{5x} - 2u_{xx} u_{4x} - 9 u^2 u_{4x} - 90 u^5 &- 120 u^3 u_{xx} + 120 u^2 u_x^2 +{}\\
&+ 48 u u_x u_{xxx} - 24 u u_{xx}^2 + 42 u_x^2 u_{xx} + u_{xxx}^2)\,.
\end{aligned}
\end{align*}
(Here we have denoted $u_{4x} = u_{xxxx}$, etc.) Since $P_1$, $P_2$ and $\varphi$ do not depend on $t$ and $x$, we get $\vartheta = \int_0^1 {G[\tau u]}/{\tau}\, d\tau$, and
\begin{align*}
\vartheta =\ & c_0(u_{4x} + 10uu_{xx} + 5u_x^2 + 10u^3) + c_1(2uu_{4x} - 2u_xu_{xxx} + u_{xx}^2 + 20u^2u_{xx} + 15u^4) {}\\
&\begin{aligned}
+\, c_2\Big(u_x u_{5x} - u_{xx} u_{4x} - 3 u^2 u_{4x} - 18 u^5 - 30 u^3 u_{xx} + 30 u^2 u_x^2 + 14 u_x^2 u_{xx} &- 8 u u_{xx}^2 +{}\\
&+ 16 u u_x u_{xxx} + \dfrac{u_{xxx}^2}{2}\Big)
\end{aligned}
\end{align*}
is the resulting constant of $X$-invariant motion. One can eliminate $u_{5x}$ using the constraint $u_{5x} = -10uu_{xxx} - 20u_xu_{xx} - 30u^2u_x$.
We consequently obtain three functionally independent constants of $X$-invariant motion implying that on $X$-invariant solutions,
\begin{align}
\begin{aligned}
&u_{4x} + 10uu_{xx} + 5u_x^2 + 10u^3 = C_0\,,\\
&2u_x u_{xxx} - u_{xx}^2 + 10uu_x^2 + 5u^4 - 2C_0u = C_1\,,\\
&\dfrac{1}{2}u_{xxx}^2 + 6 u u_x u_{xxx} + 2u u_{xx}^2 - (u_x^2 - 10 u^3 + C_0) u_{xx} + 15u^2u_x^2 + 12u^5 - 3C_0 u^2 = C_2\,,
\end{aligned}
\label{KdVconstants}
\end{align}
where $C_0, C_1, C_2\in \mathbb{R}$. One can use them together with the symmetries $\partial_t$, $\partial_x$ to completely integrate the system for $X$-invariant solutions consisting of the KdV and the equation $\varphi = 0$.

\bigskip

\remarka{The conservation law with the cosymmetry $u_{4x} + 10uu_{xx} + 5u_x^2 + 10u^3$
is also $X$-invariant. The corresponding constant of $X$-invariant motion reduces to
\begin{align*}
\dfrac{1}{2}(u_{4x} + 10uu_{xx} + 5u_x^2 + 10u^3)^2\,,
\end{align*}
which coincides with $C_0^{\hspace{0.2ex} 2}/2$.}

\subsubsection{The local general solution}

The constants of $X$-invariant motion~\eqref{KdVconstants} allow one to
express the variables $u_{4x}$, $u_{xxx}$, and $u_x$ as functions of $u$, $u_{xx}$, and $C_0$, $C_1$, $C_2$ in a neighborhood of an appropriate point. Using
\begin{align}
u_{xxx} = \dfrac{1}{2u_x}(u_{xx}^2 - 10uu_x^2 - 5u^4 + 2C_0u + C_1)
\label{thirdder}
\end{align}
and the third relation in \eqref{KdVconstants}, we get the biquadratic equation for $u_x$:
\begin{align}
au_x^4 + bu_x^2 + c = 0\,,
\label{biquadratic}
\end{align}
where the coefficients are given by
\begin{align*}
&a = -\dfrac{5}{2}u^2 - u_{xx}\,,\qquad b = \dfrac{5}{2} u u_{xx}^2 + (10 u^3 - C_0)u_{xx} - 2C_0u^2 + \dfrac{1}{2} C_1 u + \dfrac{19}{2} u^5 - C_2\,,\\
&c = \dfrac{1}{8} u_{xx}^4 + \Big(\dfrac{1}{4} C_1 - \dfrac{5}{4} u^4 + \dfrac{1}{2} C_0 u \Big) u_{xx}^2 + \dfrac{1}{2} C_0^2 u^2 - \dfrac{5}{2}C_0 u^5 -\dfrac{5}{4} C_1 u^4 + \dfrac{25}{8} u^8 + \dfrac{1}{8}C_1^2 + \dfrac{1}{2} C_0 C_1 u\,.
\end{align*}
For a certain set of $u$, $u_{xx}$, $C_0$, $C_1$, $C_2$, the equation~\eqref{biquadratic} yields real solutions $u_x$.
For instance, in a neighborhood of $u = 1$, $u_{xx} = 0$, $C_0 = C_1 = C_2 = 0$, one can choose the root
$$
u_x = \sqrt{\dfrac{-b + \sqrt{b^2 - 4ac}}{2}}\,.
$$
This formula for $u_x$ and the relations
\begin{align*}
&u_t = 6uu_x + \dfrac{1}{2u_x}(u_{xx}^2 - 10uu_x^2 - 5u^4 + 2C_0u + C_1)\,,\\
&u_{txx} = -2u_x u_{xx} - \dfrac{2u}{u_x}(u_{xx}^2 - 10uu_x^2 - 5u^4 + 2C_0u + C_1) - 30u^2 u_x\,,\\
&u_{xxx} = \dfrac{1}{2u_x}(u_{xx}^2 - 10uu_x^2 - 5u^4 + 2C_0u + C_1)
\end{align*}
allow us to replace $u_t$, $u_x$, $u_{txx}$, and $u_{xxx}$ with their expressions in terms of $u, u_{xx}, C_0, C_1, C_2$ in
\begin{align*}
du - u_t dt - u_x dx = 0\,,\qquad
du_{xx} - u_{txx} dt - u_{xxx} dx = 0
\end{align*}
or in the equivalent system (in a neighborhood of $u = 1$, $u_{xx} = 0$, $C_0 = C_1 = C_2 = 0$)
\begin{align*}
dt = \dfrac{u_{xxx}du - u_x du_{xx}}{u_t u_{xxx} - u_x u_{txx}}\,,\qquad
dx = \dfrac{-u_{txx}du + u_t du_{xx}}{u_t u_{xxx} - u_x u_{txx}}\,.
\end{align*}
It remains to integrate these two equations using Green's theorem. Denote by $A_0$, $A_2$, $B_0$, $B_2$ the corresponding functions of $u, u_{xx}, C_0, C_1, C_2$:
\begin{align*}
\dfrac{u_{xxx}}{u_t u_{xxx} - u_x u_{txx}} = A_0\,,\qquad \dfrac{ - u_x}{u_t u_{xxx} - u_x u_{txx}} = A_2\,,\\[2ex]
\dfrac{-u_{txx}}{u_t u_{xxx} - u_x u_{txx}} = B_0\,,\qquad \dfrac{u_t}{u_t u_{xxx} - u_x u_{txx}} = B_2\,.
\end{align*}
Then Green's theorem allows us to write the (local) general solution in the implicit form
\begin{align*}
&t = \int_1^u A_0(s, 0, C_0, C_1, C_2)ds + \int_0^{u_{xx}} A_2(u, s, C_0, C_1, C_2)ds + C_3\,,\\
&x = \int_1^u B_0(s, 0, C_0, C_1, C_2)ds + \int_0^{u_{xx}} B_2(u, s, C_0, C_1, C_2)ds + C_4\,.
\end{align*}
The implicit function theorem shows that locally, these relations can be rewritten in the form
\begin{align*}
u = U(t - C_3, x - C_4, C_0, C_1, C_2)\,,\qquad u_{xx} = \widetilde{U}(t - C_3, x - C_4, C_0, C_1, C_2)\,.
\end{align*}
Here $U$ is the desired general solution.

\subsection{\label{Example2} Potential Kaup-Boussinesq system}

Consider the following system of equations
\begin{align}
&v_t = -\dfrac{v_x^2}{2} - \eta_x\,,\qquad \eta_t = -v_x\eta_x - \dfrac{1}{4}v_{xxx}\,.
\label{KBClebsch}
\end{align}
Here $u^1 = v$, $u^2 = \eta$.
This system is a two-dimensional covering~\cite{VinKr} of the Kaup-Boussinesq equations
$$u_t + uu_x + h_x = 0, \qquad h_t + (hu)_x + \dfrac{1}{4}u_{xxx} = 0.
$$
The covering \eqref{KBClebsch} is determined by the Clebsch potentials $(v, \eta)$ satisfying $u = v_x$, $h = \eta_x$.

Let $X$ be the evolutionary symmetry with the characteristic $\varphi = (\varphi^1, \varphi^2)^T$, where
\beq\label{eq:KBpot:symm}
\varphi^1 = \dfrac{1}{3}v_{xxx} + 2v_x\eta_x + \dfrac{1}{3}v_x^3\,,\qquad \varphi^2 = \dfrac{1}{3}\eta_{xxx} + \dfrac{1}{2}v_x v_{xxx} + \dfrac{1}{4}v_{xx}^2 + v_x^2\eta_x + \eta_x^2\,.
\eeq
The conservation law corresponding to the cosymmetry $\psi = (\psi_1, \psi_2)$
\beq\label{eq:KBpot:CL}
\psi_1 = -c_0\eta_{xx} - c_1(v_{4x} + 4\eta_{xx} v_x + 4\eta_{x} v_{xx})\,,\qquad \psi_2 = -c_0v_{xx} - 4c_1(\eta_{xx} + v_xv_{xx})
\eeq
(here $c_0, c_1\in \mathbb{R}$ are arbitrary) is $X$-invariant because $X(\psi) + l_{\varphi}^{\hspace{0.2ex} *}(\psi) = 0$. It is represented by the differential form $P_1 dx - P_2\hspace{0.15ex} dt$, where
\begin{align*}
&P_1 = c_0v_x\eta_x + c_1\Big(2(\eta_x^2 + v_x^2\eta_x) + \dfrac{1}{2}v_x v_{xxx}\Big),
\end{align*}
while $P_2$ does not depend on $t$ and $x$. Then
\begin{align*}
X(P_1) =\ &c_0(v_x D_{x}(\varphi^2) + \eta_x D_{x}(\varphi^1)) +{}\\
&+ c_1\Big((4\eta_x + 2v_x^2) D_{x}(\varphi^2) + \Big(4v_x\eta_x + \dfrac{1}{2}v_{xxx}\Big) D_{x}(\varphi^1) + \dfrac{1}{2}v_x D_{x}^{\hspace{0.1ex} 3}(\varphi^1)\Big)\,,
\end{align*}
and
\begin{align*}
G[u] =\ & c_0\Big(\dfrac{3}{2}v_x^2v_{xxx} + 6v_x\eta_x^2 + 4v_x^3\eta_x + \dfrac{2}{3}v_x\eta_{xxx} - \dfrac{2}{3}v_{xx}\eta_{xx} + \dfrac{2}{3}v_{xxx}\eta_{x}\Big) +{}\\
&\begin{aligned}
+\, c_1\Big(\dfrac{1}{3}v_{xxx}^2 &+ \dfrac{1}{3}v_x v_{5x} - \dfrac{1}{3}v_{xx} v_{4x} - v_x v_{xx}\eta_{xx} + 13v_x v_{xxx}\eta_x + 6v_x^3v_{xxx} +{}\\
&+ 2v_x^2v_{xx}^2 + 5v_x^2\eta_{xxx} + 10v_x^4\eta_x + 32v_x^2\eta_x^2 + v_{xx}^2\eta_x + \dfrac{8}{3}\eta_x \eta_{xxx} - \dfrac{4}{3}\eta_{xx}^2 + 8\eta_x^3\Big).
\end{aligned}
\end{align*}
Here again we have denoted $v_{4x} = v_{xxxx}$, etc. Because $P_1$, $P_2$ and $\varphi$ do not depend on $t$ and $x$, we get $\vartheta = \int_0^1 {G[\tau u]}/{\tau}\, d\tau$, and
\begin{align*}
\vartheta =\ & c_0\Big(\dfrac{1}{2}v_x^2v_{xxx} + 2v_x\eta_x^2 + v_x^3\eta_x + \dfrac{1}{3}v_x\eta_{xxx} - \dfrac{1}{3}v_{xx}\eta_{xx} + \dfrac{1}{3}v_{xxx}\eta_{x}\Big) +{}\\
&\begin{aligned}
+\, c_1\Big(\dfrac{1}{6}v_{xxx}^2 &+ \dfrac{1}{6}v_x v_{5x} - \dfrac{1}{6}v_{xx} v_{4x} - \dfrac{1}{3}v_x v_{xx}\eta_{xx} + \dfrac{13}{3}v_x v_{xxx}\eta_x + \dfrac{3}{2}v_x^3v_{xxx} +{}\\
&+ \dfrac{1}{2}v_x^2v_{xx}^2 + \dfrac{5}{3}v_x^2\eta_{xxx} + 2v_x^4\eta_x + 8v_x^2\eta_x^2 + \dfrac{1}{3}v_{xx}^2\eta_x + \dfrac{4}{3}\eta_x \eta_{xxx} - \dfrac{2}{3}\eta_{xx}^2 + \dfrac{8}{3}\eta_x^3\Big)
\end{aligned}
\end{align*}
is the resulting constant of $X$-invariant motion. One can eliminate $v_{xxx}, v_{4x}, v_{5x}, \eta_{xxx}$ using the constraints $\varphi^1 = 0$, $\varphi^2 = 0$ and their differential consequences.
Thus we obtain two independent constants of $X$-invariant motion implying that on an $X$-invariant solution,
\begin{align*}
&\dfrac{1}{3}v_{xx}\eta_{xx} + v_x\eta_x^2 + \dfrac{1}{3}v_x^3\eta_x + \dfrac{1}{4}v_x v_{xx}^2 = C_0\,,\\[2ex]
&\dfrac{2}{3}\eta_{xx}^2 + \dfrac{4}{3}\eta_{x}^3 - \dfrac{1}{6}v_x^6 + \dfrac{1}{2}v_x^2 v_{xx}^2 - \dfrac{2}{3}v_x^4\eta_x - \dfrac{1}{3}v_{xx}^2\eta_x + \dfrac{4}{3}v_x v_{xx}\eta_{xx} = C_1\,,
\end{align*}
where $C_0, C_1 \in \mathbb{R}$.

\subsection{Potential Boussinesq system}

Let us demonstrate the approach based on~\eqref{charcons}. Consider the potential Boussinesq system
\beq\label{eq:pot:B}
u_t = v_x\,,\qquad v_t = \dfrac{8}{3}uu_x + \dfrac{1}{3}u_{xxx}\,.
\eeq
Here $u^1 = u$, $u^2 = v$. This system is the potential system (one-dimensional covering) for the Boussinesq equation
$$u_{tt} = \left(\dfrac{8}{3}uu_x + \dfrac{1}{3}u_{xxx}\right)_x.
$$

Let $X$ be the evolutionary symmetry of \eqref{eq:pot:B} with the characteristic $\varphi = (\varphi^1, \varphi^2)^T$,
\beq\label{eq:pot:B:symm}
\varphi^1 = v_{xxx} + 4(u_x v + uv_x)\,,\qquad \varphi^2 = \dfrac{1}{3}u_{5x} + 4uu_{xxx} + 8u_xu_{xx} + \dfrac{32}{3}u^2u_x + 4vv_x\,.
\eeq
The conservation law with the cosymmetry $\psi = (\psi_1, \psi_2)$
\beq\label{eq:pot:B:cosymm}
\psi_1 = c_1\,,\qquad \psi_2 = c_2
\eeq
(here $c_1, c_2 \in \mathbb{R}$ are arbitrary) is $X$-invariant since $X(\psi) + l_{\varphi}^{\hspace{0.2ex} *}(\psi) = 0$. Accordingly, $c_1\varphi^1 + c_2\varphi^2$ leads
to a constant of $X$-invariant motion. One can see that
\begin{align*}
c_1\varphi^1 + c_2\varphi^2 &= c_1(v_{xxx} + 4(u_x v + uv_x)) + c_2\Big(\dfrac{1}{3}u_{5x} + 4uu_{xxx} + 8u_xu_{xx} + \dfrac{32}{3}u^2u_x + 4vv_x\Big) \\
&= D_x \Big(c_1(v_{xx} + 4uv) + c_2\Big(\dfrac{1}{3}u_{4x} + 4uu_{xx} + 2u_x^2 + \dfrac{32}{9}u^3 + 2v^2\Big)\Big).
\end{align*}
Then there is a function $q(t)$ such that the derivative
\begin{align*}
D_t \Big(c_1(v_{xx} + 4uv) + c_2\Big(\dfrac{1}{3}u_{4x} + 4uu_{xx} + 2u_x^2 + \dfrac{32}{9}u^3 + 2v^2\Big) + q(t)\Big)
\end{align*}
vanishes on the system $\mathcal{E}_X$. The equations of the system \eqref{eq:pot:B} do not involve $t$, therefore we can take $q(t) = 0$, and the resulting constant of $X$-invariant motion is given by
\begin{align*}
c_1(v_{xx} + 4uv) + c_2\Big(\dfrac{1}{3}u_{4x} + 4uu_{xx} + 2u_x^2 + \dfrac{32}{9}u^3 + 2v^2\Big)\,.
\end{align*}
It follows that for each $X$-invariant solution, there are constants $C_1, C_2\in \mathbb{R}$ such that the relations
\begin{align*}
&v_{xx} + 4uv = C_1\,,\\
&\dfrac{1}{3}u_{4x} + 4uu_{xx} + 2u_x^2 + \dfrac{32}{9}u^3 + 2v^2 = C_2
\end{align*}
hold on the solution.

\section{Discussion}\label{sec:Disc}

In this work it has been shown that for a given PDE system, its conservation laws that are invariant with respect to a given Lie point, contact, or generally, local symmetry can be used to yield constants of motion holding for the corresponding symmetry-invariant solutions. These constants of motion can be systematically computed following the procedure outlined in Section~\ref{Sec2}. In particular, each constant of motion is given by explicit formula \eqref{constant}.

Compared to the procedures outlined in \cite{sjoberg2007double, sjoberg2009double, bokhari2010generalization, anco2020symmetry}, the current algorithm is more broadly applicable, in particular, in situations where the group of transformations generated by a point or contact infinitesimal symmetry is complex to work with, as well as in the general case of a higher local symmetry of the given system. In Sections \ref{Sec2} and \ref{SecEx}, examples are considered where constants of invariant motion are found for a point symmetry \eqref{eq:Burg:symm:Y} and the conservation law \eqref{BurgCL} of the Burgers equation \eqref{Burgers}, the higher symmetry \eqref{eq:KdV:5sym} and a conservation law \eqref{eq:KdV:CL} for the KdV equation \eqref{eq:KdV}, the potential Kaup-Boussinesq system \eqref{KBClebsch} with a local symmetry given by the components \eqref{eq:KBpot:symm} and a local conservation law with characteristic \eqref{eq:KBpot:CL}, and a potential Boussinesq system \eqref{eq:pot:B} with a local symmetry given by its components \eqref{eq:pot:B:symm} and a conservation law with characteristic \eqref{eq:pot:B:cosymm}.

From the computational point of view, the procedure presented in this work is fully algorithmic. An implementation of the example for the Kaup-Boussinesq potential system in \verb|Maple| software is presented in Appendix A.

We note that computations of local symmetries, including contact, point and higher symmetries, and local conservation law computations based on characteristics, can be performed with \verb|GeM| package for \verb|Maple| \cite{cheviakov2007gem, cheviakov2010computation, cheviakov2010symbolic,cheviakov2017symbolic}.

In a forthcoming paper, we will address the general case of PDE systems with $n \geq 2$ independent variables and describe the reduction mechanism for other terms on the first page of Vinogradov's $\mathcal{C}$-spectral sequence.

\section*{Acknowledgments }
The authors are grateful to NSERC of Canada for support through a Discovery grant RGPIN-2024-04308. K.D. is thankful to the Pacific Institute for Mathematical Sciences for support through a PIMS Postdoctoral Fellowship.

\section*{Data availability}
Data sharing is not applicable to this article as no new data were created or analyzed in this study.

\section*{Conflict of interest} The authors have no relevant financial or non-financial interests to disclose.

\appendix
\section{Maple code}\label{App:A}

We provide \verb|Maple| code for Example~\ref{Example2}. This code can be applied to any $1+1$ system consisting of two or one evolution equations, and works for arbitrary point, contact, and higher symmetries. A simple modification of the code allows one to consider systems of three or more equations.

\begin{verbatim}
restart;
K := 10:
\end{verbatim}
Here \verb|K| is a sufficiently large integer (exceeding orders of all derivatives that can appear).

In fact, we work only with functions from $\mathcal{F}(\mathcal{E})$. Then we consider the following restriction of the total derivative $D_x$

\begin{verbatim}
D_x := f -> diff(f, x) + add(diff(f, u1[i])*u1[i+1], i = 0 .. K+1)
+ add(diff(f, u2[i])*u2[i+1], i = 0 .. K+1):
\end{verbatim}
Here \verb|u1[i]| is the $i$-th order $x$-derivative of $u^1$, \verb|u2[i]| is the $i$-th order $x$-derivative of $u^2$. Note that $i = 0, \ldots$, i.e., $u^1$ is \verb|u1[0]|, and $u^2$ is \verb|u2[0]|. The r.h.s. of system~\eqref{KBClebsch} takes the form
\begin{verbatim}
f1 := -u1[1]^2/2 - u2[1]:
f2 := -u1[1]*u2[1] - u1[3]/4:
\end{verbatim}
Now we restrict the derivatives $u^1_{t}, u^2_t, u^1_{tx}, u^2_{tx}, u^1_{txx}, u^2_{txx}, \ldots$ to the corresponding system $\mathcal{E}$.
\begin{verbatim}
u1_t[0] := f1:
u2_t[0] := f2:
for k from 1 by 1 to K+1 do
    u1_t[k] := D_x(u1_t[k-1]):
    u2_t[k] := D_x(u2_t[k-1]):
end do:
\end{verbatim}
This allows us to introduce the total derivative $\,\overline{\!D}_t$
\begin{verbatim}
D_t := f -> diff(f, t) + add(diff(f, u1[i])*u1_t[i], i = 0 .. K+1)
+ add(diff(f, u2[i])*u2_t[i], i = 0 .. K+1):
\end{verbatim}
The characteristic $\varphi$ and total $x$-derivatives of its components are given by
\begin{verbatim}
phi1[0] := u1[3]/3 + 2*u1[1]*u2[1] + u1[1]^3/3:
phi2[0] := u2[3]/3 + u1[1]*u1[3]/2 + u1[2]^2/4 + u1[1]^2*u2[1] + u2[1]^2:
for k from 1 by 1 to K+1 do
    phi1[k] := D_x(phi1[k-1]):
    phi2[k] := D_x(phi2[k-1]):
end do:
\end{verbatim}
The corresponding evolutionary symmetry $X = E_{\varphi}$ is
\begin{verbatim}
Evolutionary_derivative := f -> add(diff(f, u1[i])*phi1[i], i = 0 .. K+1)
+ add(diff(f, u2[i])*phi2[i], i = 0 .. K+1):
\end{verbatim}
Now we check that this is indeed a symmetry. The outputs are supposed to be zeros.
\begin{verbatim}
simplify(D_t(phi1[0]) - Evolutionary_derivative(f1));
                               0
simplify(D_t(phi2[0]) - Evolutionary_derivative(f2));
                               0
\end{verbatim}
Both components $P_1$, $P_2$ of the conservation law representative $P_1dx - P_2 dt$ are required in order to make all further checks
\begin{verbatim}
P1 := c0*u1[1]*u2[1] + c1*(2*(u1[1]^2*u2[1] + u2[1]^2) + u1[1]*u1[3]/2):
P2 := c0*(u1[1]^2*u2[1] + u2[1]^2/2 - u1[2]^2/8 + u1[1]*u1[3]/4)
+ c1*(2*u1[1]^3*u2[1] + u1[3]*u2[1]
+ (u1[2]*u1_t[1] + u1[1]^2*u1[3] + u1[1]*(8*u2[1]^2 - u1_t[2]))/2):
\end{verbatim}
Check that $P_1dx - P_2 dt$ represents a conservation law. The output is supposed to be zero.
\begin{verbatim}
simplify(D_t(P1) + D_x(P2));
                               0
\end{verbatim}
The $X(P_1)$ is
\begin{verbatim}
Evolutionary_derivative_of_P1 := Evolutionary_derivative(P1):
\end{verbatim}
Check that the conservation law is $X$-invariant. The results are supposed to be zeros.
\begin{verbatim}
Variational_u1_derivative_of_Evolutionary_derivative_of_P1 :=
diff(Evolutionary_derivative_of_P1, u1[0]):
Variational_u2_derivative_of_Evolutionary_derivative_of_P1 :=
diff(Evolutionary_derivative_of_P1, u2[0]):
for k from 1 by 1 to K+1 do
    Ans1 := diff(Evolutionary_derivative_of_P1, u1[k]):
    Ans2 := diff(Evolutionary_derivative_of_P1, u2[k]):
    for j from 1 by 1 to k do
        Ans1 := D_x(Ans1):
        Ans2 := D_x(Ans2):
    end do:
    Variational_u1_derivative_of_Evolutionary_derivative_of_P1 :=
    Variational_u1_derivative_of_Evolutionary_derivative_of_P1 + (-1)^k*Ans1:
    Variational_u2_derivative_of_Evolutionary_derivative_of_P1 :=
    Variational_u2_derivative_of_Evolutionary_derivative_of_P1 + (-1)^k*Ans2:
end do:
simplify(Variational_u1_derivative_of_Evolutionary_derivative_of_P1);
                               0
simplify(Variational_u2_derivative_of_Evolutionary_derivative_of_P1);
                               0
\end{verbatim}
Find $G[u]$.
\begin{verbatim}
G := 0:
for k from 1 by 1 to K+1 do
    for j from 0 by 1 to k-1 do
        Ans1 := diff(Evolutionary_derivative_of_P1, u1[k]):
        Ans2 := diff(Evolutionary_derivative_of_P1, u2[k]):
        for s from 1 by 1 to k-1-j do
            Ans1 := D_x(Ans1):
            Ans2 := D_x(Ans2):
        end do:
        G := G + (-1)^(k-1-j)*Ans1*u1[j] + (-1)^(k-1-j)*Ans2*u2[j]:
    end do:
end do:
\end{verbatim}
Find $G[\tau u]$.
\begin{verbatim}
G_with_tau := simplify(G):
for k from 0 by 1 to K+1 do
    G_with_tau := subs(u1[k] = tau*w1[k], u2[k] = tau*w2[k], G_with_tau):
end do:
\end{verbatim}
Then find $\vartheta$ up to a function $h(t, x)$.
\begin{verbatim}
theta_of_w := int(G_with_tau/tau, tau = 0 .. 1):
for k from 0 by 1 to K+1 do
    theta_of_w := subs(w1[k] = u1[k], w2[k] = u2[k], theta_of_w):
end do:
theta_incomplete := theta_of_w + h(t, x):
\end{verbatim}
The equation for $h(t, x)$ results in
\begin{verbatim}
g1 := rhs(isolate(simplify(Evolutionary_derivative_of_P1
- D_x(theta_incomplete) = 0), diff(h(t, x), x))):
g2 := rhs(isolate(simplify(-Evolutionary_derivative(P2)
- D_t(theta_incomplete) = 0), diff(h(t, x), t))):
\end{verbatim}
Here we determine $g_1(t, x)$ and $g_2(t, x)$. Finally, the desired constant of $X$-invariant motion arises:
\begin{verbatim}
g1s := subs(t = 0, x = S, g1):
g2s := subs(t = S, g2):
theta_final := theta_incomplete - h(t, x)
+ int(g1s, S = 0 .. x) + int(g2s, S = 0 .. t);
\end{verbatim}
In this case, the result coincides with $\vartheta$ from Example~\ref{Example2}.
Now it remains to check that \verb|theta_final| is correct. The outputs of the following commands are indeed zeros:
\begin{verbatim}
simplify(Evolutionary_derivative_of_P1 - D_x(theta_final));
                               0
simplify(-Evolutionary_derivative(P2) - D_t(theta_final));
                               0
\end{verbatim}

To summarize, four types of input data depend on the particular situation:
the sufficiently large number \verb|K|, the right-hand sides \verb|f1|, \verb|f2| of a system under consideration, the components \verb|phi1[0]|, \verb|phi2[0]| of the characteristic, the components \verb|P1|, \verb|P2| of the conservation law representative. Checks are supposed to result in seven zeros.

To consider a scalar $1+1$-dimensional evolution equation in the same manner, one can simply take
\begin{verbatim}
f2 := 0:
phi2[0]:= 0:
\end{verbatim}
with no other changes necessary.

%
%
%
%
%

\bibliographystyle{ieeetr}
{\small
\bibliography{DC}
}

\end{document}